\documentclass[twocolumn,preprintnumbers,amsmath,amssymb]{revtex4}
\usepackage{epsfig}

\begin{document}
\title{Nanosecond electro-optical switching with a repetition rate above 20\,MHz.}
\author{Holger M\"uller}
\email{holgerm@stanford.edu}
\author{Sheng-wey Chiow, Sven Herrmann, and Steven Chu}
\affiliation{Physics Department, Varian
226, 382 Via Pueblo Mall, Stanford, California 94305}

\begin{abstract}
We describe an electro-optical switch based on a commercial
electro-optic modulator (modified for high-speed operation) and a
340\,V pulser having a rise time of 2.2\,ns (at 250\,V). It can
produce arbitrary pulse patterns with an average repetition rate
beyond 20\,MHz. It uses a grounded-grid triode driven by
transmitting power transistors. We discuss variations that enable
analog operation, use the step-recovery effect in bipolar
transistors, or offer other combinations of output voltage, size,
and cost.
\end{abstract}

\maketitle


\begin{figure}
\epsfig{file=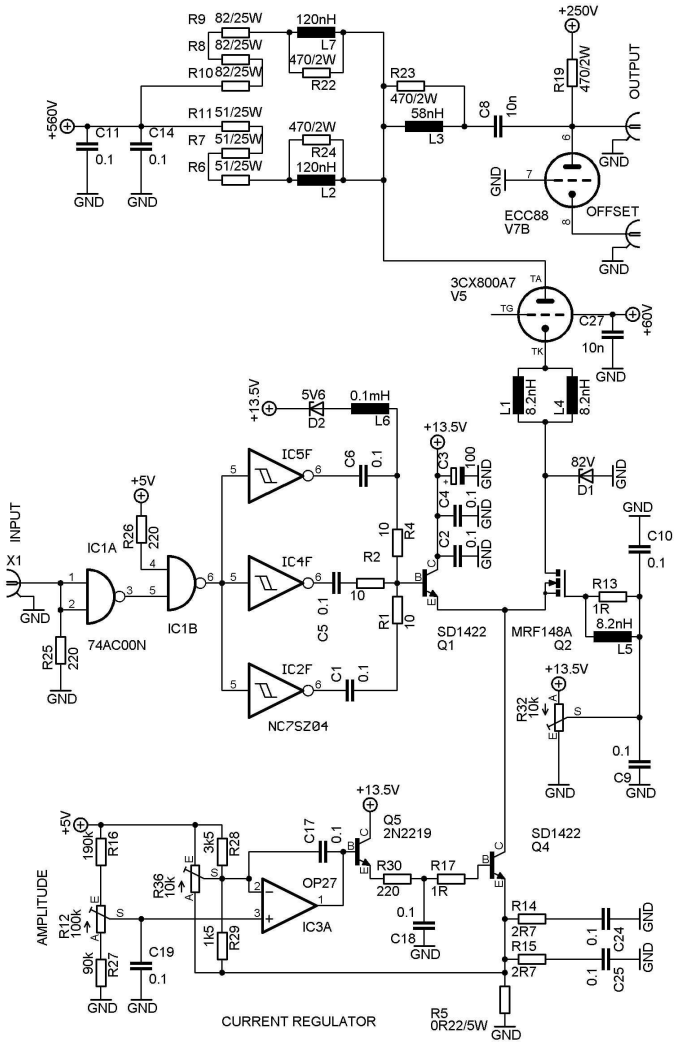,width=0.5\textwidth}
\caption{\label{schem} Schematic}
\end{figure}

We generate pulses of laser light that can have a duration below
5\,ns and a repetition rate in excess of 20\,MHz by switching a
continuous-wave laser with an electro-optic modulator (EOM). Such
pulses allow one to coherently drive atoms into the excited state
of a dipole-allowed transition and subsequently observe the decay
of the fluorescence when the laser beam is essentially switched
off and little stray light can reach the detector. This enables
low-noise fluorescence detection of atoms. For example, the $1/e$
lifetimes $\Gamma^{-1}$ are 32\,ns for Cs at a wavelength of
852\,nm and 26\,ns for Rb at 780\,nm. For coherent excitation of
essentially all the population, a flash of light having a duration
$\ll \Gamma^{-1}$ (preferably 5\,ns or lower) is required; after
an interpulse interval of, e.g., $3\Gamma^{-1}\sim 100$\,ns,
during which the population decays into the ground state, the next
pulse can be applied. This implies a repetition rate of 10\,MHz.

Our EOM approximates a capacitive load of 15\,pF and requires
about 250\,V input voltage. Nanosecond and even sub-ns rise-times
$t_r$ have been realized with avalanche transistors
\cite{Fulkerson,Henebry,Jinyuan,Rutten}, but the repetition rate
is limited to tens or hundreds of kHz by average power dissipation
and lifetime issues. Mosfets and bipolar transistors of sufficient
speed ($\sim 100\,$MHz bandwidth corresponding to $t_r=3.5$\,ns
rise-time) are limited to below 100\,V \cite{HMmodulators};
modules consisting of many devices connected in series allow
high-voltage operation, with repetition rates limited to tens of
kHz. A push-pull bipolar design capable of $t_r=1$\,ns at a
repetition rate limited only by this rise time reaches 60\,Vpp
output \cite{HMmodulators}. To surpass these limits and meet our
requirements, we use a grounded-grid vacuum triode driven by a
solid-state differential amplifier stage.\\


The input stage of the circuit shown in Fig. \ref{schem} (IC1,
IC2,IC4,IC5) receives TTL signals and restores a good square-wave
shape. The unused input of IC1b can be used by an overload
protection circuit (not shown) to disable the pulser. IC2, IC4,
IC5 provide a 5V pp pulse with about 1\,ns rise-time. AC coupling
helps to protect the power stage against excessive duty cycle.

\begin{figure}
\centering \epsfig{file=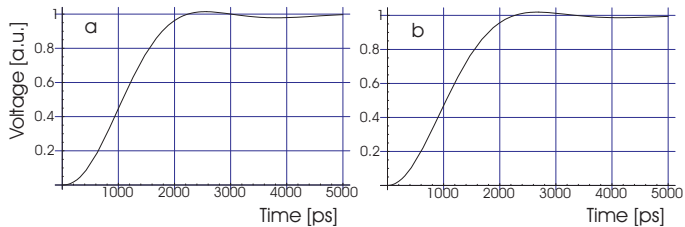,width=0.5\textwidth}
\caption{\label{pulsenorm} a: Calculated output pulse for an ideal
rectangular drive of V5. b: same for V5's input circuit, assuming
16.6$\,\Omega$ input resistance and 12\,nH lead inductance of the
tube.}
\end{figure}

Q1 and Q2 form a differential amplifier. Q1 is biased via D2 and
L6, that provide about 5.3V at the base. Between pulses, Q1
conducts and Q2 is cut off. During a pulse, the current is
directed through Q2 and the output tube. As it is regulated by Q4
and IC3A, the output amplitude is precisely constant, without
feedback over the output stage. Also, the total current being
constant, the inductance of the ground connection is not critical.
This helps to obtain a very good, fast-rise square-shaped output
waveform. Q1 can be a low voltage type, which typically leads to
high transition frequency. Since the capacities of Q1 and Q4 are
not as critical, transistors capable of high power dissipation can
be chosen. The speed of this driver stage is mainly limited by the
output capacity of 35\,pF of Q2 (MRF148A), and the tube's input
capacity of 30\,pF. Inductors L1 and L4 provide series peaking
\cite{Valley} to speed up the signal delivered to the output
stage. They are connected to two cathode pins of V5 to control
stray inductance. Simulation of this circuit (Fig.
\ref{pulsenorm}, a) shows a 10 to 90\% rise-time of 1.38\,ns.

Q2 is mounted upside down on top of Q1 so that both source leads
of Q2 can be connected to Q1 to obtain low parasitic inductance
$L_{Q1Q2}$. For example, just 1\,nH would contribute about
$2.2\times L/R\sim$ 1.3\,ns to the rise-time, where $R\sim$
3\,A/5\,V is the impedance of this circuit. The Zener diode D1
protects Q2 against excessive voltage. Oscillations at 250\,MHz
are suppressed by the capacity of D1, the resistors and LR
combinations in the base leads of the transistors, and the R-C
combinations in the emitter lead of Q4.

The pulser is designed with an output impedance of $93\,\Omega$.
Coaxial cables for this are readily available, and the required
peak current is reduced to one half of that required for
50\,$\Omega$. This impedance together with the load capacity sets
a lower limit of about $t_r\sim 2$\,ns for the rise-time. We will
closely approach this theoretical limit.

The tube 3CX800A7 \cite{Eimacspecs} (Eimac division of CPI, Inc.)
is chosen primarily for its high gain and can deliver up to 8\,A
of peak anode current. It requires $\sim 60$\,V drive voltage and
0.2\,A peak grid current for 300\,V output. A special capacitor
(Eimac) having low-inductance, large-area contacts bypasses the
grid to ground. The dc grid voltage is adjustable from 20 to about
60\,V. The output signal is ac-coupled to the output via C8. V7 is
used as a clamp diode to provide a low impedance path for charging
C8, thus making the output voltage between pulses very close to
zero. Use of a vacuum triode ECC88, connected as a diode, avoids
possible pulse distortion associated with the recovery of a
semiconductor diode. To minimize stray capacities, it is soldered
directly into the circuit without a socket and its internal shield
of V7 is left floating. By removing the ground connection at the
OFFSET connector, V7 is cut off and a positive bias of 240\,V
appears at the output. If desired, a potentiometer or a variable
voltage between 0 and 10\,V at the OFFSET connector can be used to
vary the dc level at the output between about 5 and 240\,V. This
can be used for applying slow, continuous signals to the
modulator.

R7-R10 are low-inductance types of very compact design (Ohmite
TA025PW82R0JE and TA025PW51R0JE), arranged as depicted in the
circuit in a way to minimize stray capacities and inductance.
Furthermore, the output circuit consists of L2, L7 (112\,nH each)
and L3 (56\,nH) with paralleled resistors, the output capacity
$C_a=15\,$pF of the triode (including circuit strays as measured
in our setup), a capacity $C_d=4.5\,$pF of the diode, and a
capacity $C_r=4.5$\,pF of the resistors. Simulation shows a
rise-time of 1.31\,ns for the output circuit. The combined
risetime of the tube's driver and output circuits is thus
1.90\,ns. Fig. \ref{pulsenorm} shows the calculated response of
the output circuit.

\begin{figure}
\centering \epsfig{file=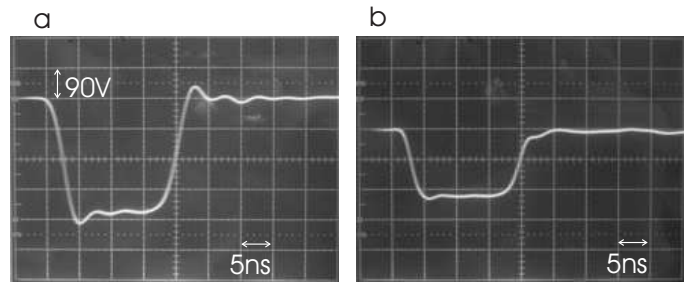,width=0.5\textwidth}
\caption{\label{pulsemeas} (a): Electrical output waveform at an
electrical amplitude of 340\,V. (b): optical waveforms using a New
Focus model 4102 EOM driven with about 250\,V via about 2\,m of
RG-63 coaxial cable.}
\end{figure}

The maximum duty cycle $\eta$ that is permissible for the 3CX800A7
is specified by the 'rms current rule,' that states that the root
mean square (rms) current $\sqrt{\eta} \hat I$ should not exceed
the maximum permissible dc current of 600\,mA \cite{Eimacspecs}.
However, when operating the tube as a class B or C radio frequency
amplifier, rms currents 1.6-2 times higher are permissible. Using
this as the real limit on the rms current, the permissible duty
cycle rises by a factor of 2.7-4. Indeed, the data sheets of some
tubes specify duty cycles exceeding the rms current rule (e.g.,
E130L, 8533, 8906). Thus, operation at about 10\% duty should be
quite safe. With a minimum pulse duration (limited by the rise
time) of below 4\,ns full width at half maximum, this sets a limit
on the repetition rate in excess of 20\,MHz.

At 10\% duty at $f$=20\,MHz and 3.2A peak, the plate of V5
dissipates $<180$\,W including $\sim 50$\,W switching losses
assuming 15\,pF plate capacity. Q2 dissipates about 26W, including
22W switching losses (assuming that the losses when switching on
and off are the same), well within the manufacturer's
specification of 115\,W. For Q1 and Q4, losses are 48\,W (sum of
both transistors) are almost independent of duty cycl between 0
and 10\%.

The output amplitude is determined by the setting of the
constant-current source by R12. The transistor capacities and
circuit impedances change with peak current which may lead to
pulse distortion (overshoot and ringing). To prevent this, the dc
grid voltage $U_g$ of V5 is adjusted between 20-75\,V according to
the setting of the constant-current source. The optimum is
obtained for $U_g$ such that Q2 is just above saturation.

This leads to a very good square waveform having less than 5\%
overshoot. At an amplitude 250\,V, the rise-time is $t_r=2.2$\,ns.
This is what is expected from the 1.9\,ns theoretical rise-time
for Q2 and the power stage and 1\,ns due to the NC7SZ04 inverters
($\sqrt{1.9^2+1^2}=2.1$), but appears too fast in view of
additional influences like Q1's transition frequency. A possible
reason for that is that $L_{\rm Q1Q2}$ provides additional series
peaking. At the maximum amplitude of 340\,V, which is limited by
the constant current source, the rise-time increases to 2.7\,ns,
see Fig. \ref{pulsemeas} (a). This may be in part because Q2 is
then operated closer to saturation which increases its capacity
and doesn't allow its drain potential to swing as required for
series peaking by L1 and L4.\\


The EOM (New Focus model 4102) uses two lithium niobate crystals
electrically connected in parallel, which receive signals via an
internal 50$\,\Omega$ strip-line of about 2.5\,cm length from an
SMA connector. The total capacity is about 16\,pF. Directly
driving this from a $Z_0=93\,\Omega$ cable, $t_r\approx
2.2Z_0C\approx 3.3$\,ns, even if the pulser output would be
infinitely fast. As this would significantly reduce the overall
speed, we increase match the characteristic impedance of the
strip-line to 93\,$\Omega$ by reducing its width. As the crystals
are now directly driven from a 93$\,\Omega$ source, we only need
to consider the crystal capacity of about 12\,pF. The speed is
further enhanced by series peaking \cite{Valley} with an inductor
of 63\,nH in series to the crystals. The EOM should now show
$t_r\sim 1.8$\,ns. With an electrical $t_r=2.2\,$ns of the pulser,
we expect an optical $t_r\approx 0.66\times
\sqrt{2.2^2+1.8^2}$\,ns=1.9\,ns. The factor of 0.66 is the
theoretical conversion factor for a Gaussian frequency response
and the $\sin^2$ transfer function of the EOM. This agrees well
with the observed optical $t_r=2$\,ns, Fig. \ref{pulsemeas} (b).\\


The circuit can be adapted for analog operation, as a linear
amplifier, by driving at the base of Q1. If directly driven from a
a $50\,\Omega$ generator terminated into $50\,\Omega$, the speed
is limited by the capacitive ($\sim 70\,$pF) input impedance of Q1
and the finite transition frequency $f_T$ of Q1, which in
connection with the approximately 1.5$\,\Omega$ Ohmic load due to
Q2 appears as an additional effective input capacity of $\sim
100\,$pF (assuming $f_T \sim 1\,$GHz). Thus, the RC time constant
is 4.2\,ns. With a suitable input network, a rise-time of 6-8\,ns
should be possible. It can be improved by reducing the impedance,
either by a transformer or by an additional emitter follower.

Using a lower value for R5, the maximum output voltage can be
increased to 450\,V, limited by the maximum collector current of
Q1. Using larger transistors is also possible, especially as Q1's
capacity is relatively insignificant. The 3CX800A7 tube enables up
to $\sim$800\,V; if 300\,V output or less suffice, a smaller and
cheaper tube may be substituted. Using a E130L \cite{E130L} should
lead to similar performance, save for a lower maximum duty cycle.

If the electro-optic crystals and the electronics can be in the
same housing, the output impedance can be chosen independently of
cable impedances. For a EOM capacity of 12pF plus 6pF tube and
wiring capacity, with series-shunt peaking, an electrical
rise-time of 3\,ns can be achieved with $R_2=167\Omega$, requiring
only 1.5\,A peak current. A QQE06/40 tube, that could deliver such
current, has an output capacity of 6.4pF.

A previous version of this circuit used a bipolar npn 2N5642 as
Q2. Near the maximum output amplitude of 250\,V (lower because of
the voltage rating of the 2N5642), the trailing edge became slow,
about 5-10\,ns. Furthermore decreasing the grid potential of V5
eventually leads to saturation of the transistor. This increases
the pulse duration by a few ns, as expected because of the
minority carrier charge storage in the transistor. However, we
accidentally found that it also leads to a {\em speed-up} of the
trailing edge. We attribute this to the step recovery effect of
the transistor \cite{Paul,Gerding}: As soon as the stored minority
carriers are removed, the transistor switches off rapidly. Thus,
rise and fall times of 3.3 and 3.1\,ns were achieved. All 5 tested
samples of 2N5642 (made by Advanced Semiconductor) worked
satisfactorily.\\

This material is based upon work supported by the National Science
Foundation under Grant No. 0400866, the Air Force Office of
Scientific Research, and the Multi-University Research Initiative.


\begin{thebibliography}{99}

\bibitem{Fulkerson} 
E. Stephen Fulkerson, Douglas C. Norman, and Rex Booth in: Digest
of Technical Papers. 11th IEEE International Pulsed Power
Conference, 1997.

\bibitem{Henebry} W.M. Henebry, Rev. Scientific Instrum. {\bf 32}, 1198 (1961).

\bibitem{Jinyuan} L. Jinyuan, S. Bing, and C. Zenghu, Rev.
Sci. Instrum. {\bf 69,} 3066 (1998).

\bibitem{Rutten} T.P. Rutten, N. Wild, and P.J. Veitch, Rev. Sci.
Instr. {\bf 78,} 073108 (2007).


\bibitem{HMmodulators} H. M\"uller, Rev. Scientific Instrum. {\bf 76}, 084701 (2005).


\bibitem{Valley} R.M. Walker and H. Wallman in: G.E. Valley and H. Wallman (Eds.), {\em Vacuum-Tube Amplifiers} (McGraw-Hill, New York, 1948).

\bibitem{Eimacspecs} 3CX800A7 data sheet (Eimac division of CPI, Inc, San Carlos, CA 1987).

\bibitem{E130L} E130L special quality high slope power pentode data sheet (Philips, Eindhoven, 1961) (see, e.g., http://frank.pocnet.net/index.html.).

\bibitem{Paul} R. Paul, {\em Transistoren} (Vieweg, Braunschweig, 1965).

\bibitem{Gerding} M. Gerding, T. Musch, and B. Schiek, Advances in Radio Science {\bf 2}, 7-12 (2004).



\end{thebibliography}
\end{document}